\documentclass{ws-acs}

\begin{document}

\markboth{Dami\'{a}n H. Zanette} {Self-similarity in the taxonomic
classification of human languages}

\catchline

\title{SELF-SIMILARITY IN THE TAXONOMIC CLASSIFICATION OF HUMAN LANGUAGES}

\author{\footnotesize DAMIAN H. ZANETTE\footnote{email: zanette@cab.cnea.gov.ar}}

\address{Consejo Nacional de Investigaciones Cient\'{\i}ficas y
T\'{e}cnicas\\Centro At\'{o}mico Bariloche and Instituto Balseiro\\
8400 Bariloche, R\'{\i}o Negro, Argentina }

\maketitle

\pub{Received (received date)}{Revised (revised date)}

\begin{abstract}
Statistical properties of the taxonomic classification of human
languages are studied. It is shown that, at the highest levels of
the taxonomic hierarchy, the frequency of taxon members as a
function of the number of languages belonging to each member
decays as a power law. This feature reveals that a self-similar
structure underlies the taxonomy of languages, exactly as
observed in the taxonomic classification of biological species.
Such an analogy is a clue to the evolutionary foundation of
language classification based on long-range comparison.
\end{abstract}

\section{Introduction}

Comparative linguistics shows that human languages can be grouped
in a hierarchy of families whose members share a certain level of
similarity, much like biological species in the taxonomic tree.
This hierarchy is defined in terms of mutual relatedness and
affinity of languages in their present form, but also takes into
account evolutionary aspects such as common innovations. Ruhlen
\cite{ruhlen87} compiled data for the almost 5,000 extant
languages and proposed a taxonomic classification which, at the
highest level, consists of 17 families. Some instances of these
large families are the Indo-Hittite, which contains all
Indo-European languages and is the largest in number of speakers;
the Austric, which covers parts of South-Eastern Asia and Oceania
and is the richest in number of languages; and the Amerind, which
was one of the latest to be recognized as a family
\cite{greenberg87}. These families are divided into primary
branches, which in turn contain groups, subgroups, branches, and
so on. Along certain particularly rich branches (e.g. Bantu, in
Africa) Ruhlen's classification distinguishes up to 17
hierarchical levels or taxa.

The methods of long-range comparison that make possible the
identification of language families at the highest taxonomic
levels have been emphatically criticized by many linguists
\cite{kaufman90,hock91}. These authors claim that an upper bound
of 6,000-8,000 years exists for the time elapsed from the
separation of two languages from a common ancestor such that any
connection between them can be established by comparison.
Ruhlen's work, however, has received strong support from a field
outside linguistics, namely genetics, through detailed studies of
the genetic distance between human populations by Cavalli-Sforza
and coworkers \cite{cavalli88}. These authors have shown that, at
the highest levels, the taxonomy of populations is remarkably
similar to that of languages. The similarity can be traced up to
levels corresponding to the main population expansions towards
Eastern Asia, Oceania, and the Americas, some tenths of thousand
years ago \cite{ruhlen94}.

In this paper, statistical regularities of the taxonomic
classification of languages at the highest taxa are disclosed.
The hierarchical distribution of languages is shown to exhibit
self-similarity properties, which could hardly be explained if
such distribution were derived from a baseless method. Comparison
with the case of biological species, in fact, supports an
evolutionary basis for the classification of languages.

\section{Analysis}

Our statistical analysis proceeds as follows. We choose a
specific taxon of the hierarchy (say, primary branches). For each
member $i$ of that taxon (say, Indo-European) we determine the
number $n_i$ of extant languages belonging to that member (for
Indo-European, $n_i = 144$).  The set of values $n_i$ obtained for
the selected taxon is then used to construct a histogram. The
height of each column in the histogram is proportional to the
fraction of members whose number of languages lies within the
interval covered by the column, normalized by the column width.
In other words, it gives the frequency $f(n)$ of  taxon members
which contain a given number $n$ of languages.

\begin{table}[h]
\tbl{Statistical parameters of the taxonomic classification. The
exponents $\gamma$ and $\nu$ characterize the power-law dependence
of the frequency of members of a given taxon on the number of
languages and the number of members in the successive taxon,
respectively. The regression coefficient measures the quality of
the least-square fitting from which $\gamma$ is obtained. }
{\begin{tabular}{@{}cccc@{}} \toprule taxon & exponent $\gamma$ &
regression coefficient & exponent $\nu$ \\  \colrule
first &  $1.0 \pm 0.2$ & $-0.903$ &   $1.0 \pm 0.2$\\
second & $1.4 \pm 0.1$ & $-0.976$ &   $1.7 \pm 0.1$\\
third &  $1.6 \pm 0.1$ & $-0.990$ &   $1.7 \pm 0.1$ \\
fourth & $1.9 \pm 0.1$ & $-0.993$ &   $1.9 \pm 0.1$ \\
fifth &  $2.1 \pm 0.1$ & $-0.998$ &   $-$ \\ \botrule
\end{tabular}}
\label{t1}
\end{table}

\begin{figure}[h]
\centerline{\psfig{file=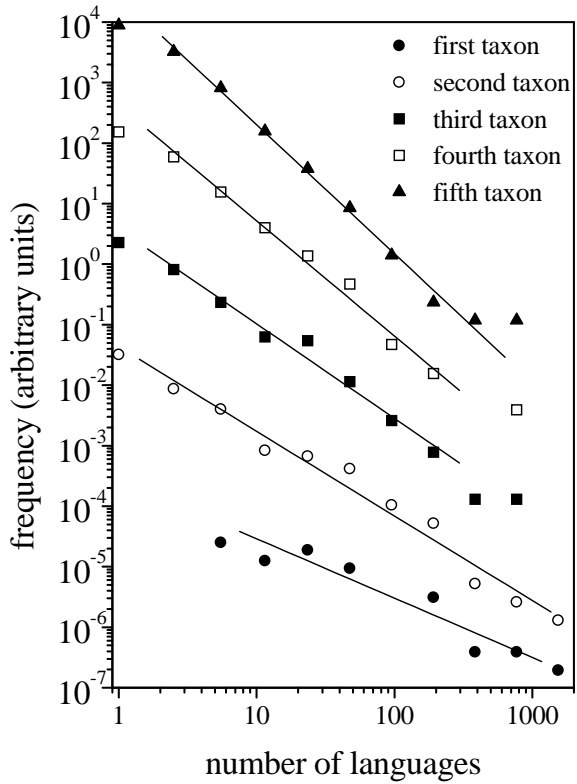,width=9cm,angle=0}}
\vspace*{8pt}
\caption{Frequency of taxon members, shown as a function of the
number of languages belonging to each member, for the first five
taxonomic levels  (families, primary branches, groups, subgroups,
and branches).  For clarity in displaying, the frequencies of
each set have been multiplied by an appropriate constant. The
lines correspond to least-square fittings in the intervals where
they are plotted.}
\label{f1}
\end{figure}

Results for the highest, first five taxa (families, primary
branches, groups, subgroups, and branches) are presented in Fig.
\ref{f1}. For clarity, the histograms are displayed as sets of
points, and the frequencies corresponding to each set are
expressed in arbitrary units. In all cases, the data exhibit a
regime of well-defined power-law decay, $f(n) \sim n^{-\gamma}$,
spanning more than two decades in the number of languages,
typically from $n \approx 2$ to $n \approx 300$, and three to
four decades in frequencies. The exponent $\gamma$, obtained from
linear least-square fitting on the log-log plot, is given for
each set in Table \ref{t1}. The fittings are shown in Fig.
\ref{f1} as straight lines. As a measure of the fitting quality,
the regression coefficients are also given in Table \ref{t1}. They
are always above $0.9$ (in modulus).

\vspace{10 pt}

\begin{figure}[h]
\centerline{\psfig{file=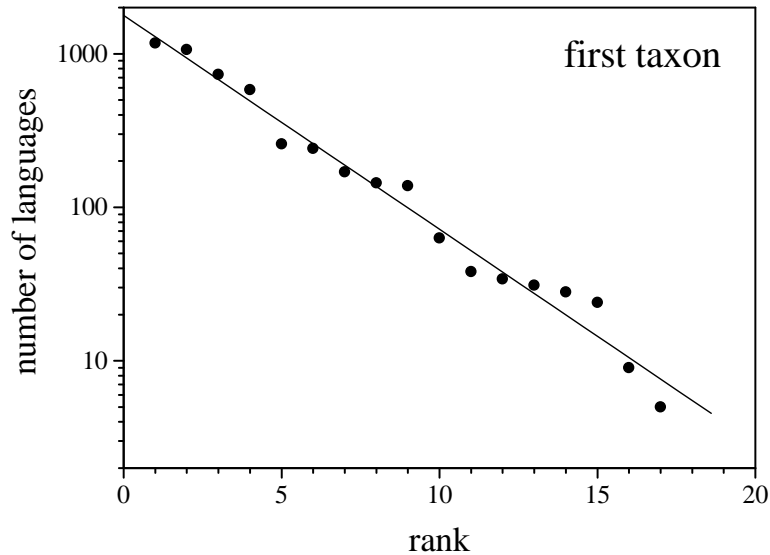,width=12cm,angle=0}}
\vspace*{8pt}
\caption{Rank plot of the first taxon (families). The number
of languages belonging to each member is plotted as a function
of its rank. The straight line corresponds to an exponential
decay and has been determined by least-square fitting.}
\label{f2}
\end{figure}

The regression coefficients show that the definiteness of the
power-law dependence improves for lower taxa. This is also
apparent from Fig. \ref{f1}, where the linear approximation is
relatively poorer for the first taxa. We ascribe this effect to
the fact that the number of members of a given taxon decreases
considerably as higher taxa are considered. For the first taxon,
in fact, the ($8$-column) histogram is constructed from a set of
only $17$ values. In this specific case, it is more reliable to
study the distribution of languages in families using a rank plot
\cite{zipf49}. The rank $r$ of a family is given by its place in
a list where families are sorted in decreasing order by the
number of languages belonging to them ($r = 1$ for the richest
family, $r = 2$ for the second richest, and so on). The rank plot
displays the number of languages as a function of $r$, as shown
in Fig. \ref{f2}. In this linear-log plot the straight line stands
for an exponential decay, $n \sim \exp(-a r)$, and can be shown to
correspond to a frequency of the form $f(n) \sim n^{-1}$. These
data are reasonably well approximated by a linear fit (regression
coefficient $= -0.988$), which is in full agreement with the
corresponding power-law exponent $\gamma = 1.0 \pm 0.2$, quoted
in Table \ref{t1}.

The fact that the frequency $f(n)$ exhibits power-law dependence
for several taxa implies another important statistical property of
the taxonomic classification. Consider two consecutive taxa $t$
and $t+1$ with frequencies $f_t(n) \sim n^{-\gamma}$ and
$f_{t+1}(n) \sim n^{-\gamma'}$, respectively, and call $p(m)$ the
fraction of members of taxon $t$ that contain a number $m$ of
members of the successive taxon $t+1$. These three distributions
are related through the expression
\begin{equation}
f_t(n) = \sum_m p(m)  f_{t+1}(n) \circ f_{t+1}(n) \circ \cdots
\circ f_{t+1}(n),
\end{equation}
where the $m$-th term involves the $m$-fold discrete convolution
of $f_{t+1}$ with itself. In the Laplace domain, this relation
reads
\begin{equation} \label{Lap}
\phi_t(s)= \sum_m p(m) [\phi_{t+1}(s)]^m \approx \int_0^\infty
p(\mu) [\phi_{t+1}(s)]^\mu d\mu,
\end{equation}
where $\phi_{t}$ and $\phi_{t+1}$ are the Laplace transforms of
$f_{t}$ and $f_{t,t+1}$, respectively. In the right-hand side of
Eq. (\ref{Lap}), the variable $\mu$ replaces the summation index
$m$ to produce a continuous approximation to $\phi_t$. The
power-law decay of $f_t(n)$ implies that, near $s=0$, its Laplace
transform behaves as $\phi_t(s) \approx \exp(-a|s|^{\gamma-1})$
for $1<\gamma<2$ and as $\phi_t(s) \approx \exp(-bs-
c|s|^{\gamma-1})$ for $2<\gamma<3$, where $a$, $b$, and $c$ are
constant coefficients \cite{Doe}. Analogous approximate
expressions hold for $\phi_{t+1}(s)$. These asymptotic
expressions satisfy the continuous approximation in Eq.
(\ref{Lap}) if the distribution $p(m)$ is in turn a power law for
large $m$, $p(m) \sim m^{-\nu}$. The exponent $\nu$ is a function
of $\gamma$ and $\gamma'$, namely,
\begin{equation}
\nu = 1+\frac{\gamma-1}{\gamma'-1}
\end{equation}
if $1<\gamma,\gamma'<2$, and
\begin{equation}
\nu = \gamma
\end{equation}
if $1<\gamma<2<\gamma'$. Its value for the first four taxa is
also given in Table \ref{t1}.

\section{Discussion and conclusion}

Power-law frequency distributions are known to reveal
self-similarity and fractal geometry in the underlying structures
\cite{mandelbrot97} --in our case, the taxonomic tree. The
interest of this statistical property of the taxonomic
classification of languages resides in the fact that exactly the
same feature is found in the taxonomy of biological species. The
power-law dependence in the frequency of biological taxon
abundance has been first discussed by Yule \cite{yule24} and, much
later, Burlando \cite{burlando90,burlando93} studied in detail the
distribution of the exponent $\nu$ at different levels and along
different branches of the taxonomic tree, also including some
families of extinct species. In contrast with the case of
languages, the exponent $\nu$ for biological taxonomy can
directly be measured on the tree. Indeed, the biological
taxonomic tree is very rich --it contains more than $1,500,000$
species at the lowest level. Even at the highest taxa, one finds
members with a large number of members from the successive taxon.
On the other hand, the extant languages are less than $5,000$.
The statistics are consequently much poorer, and the exponent
$\nu$ is more reliably inferred from the values of $\gamma$ and
$\gamma'$, as done above. It has been found that, for biological
species, $\nu$ varies in a relatively narrow interval,
$1.4<\nu<2.5$. Note that, except for the first taxon, the values
of $\nu$ obtained for language taxonomy are also in that interval.

Several models have been proposed to account for the statistical
regularities in the taxon abundance of biological species,
ranging from branching dynamical processes
\cite{yule24,matsumoto99,chu99}  to simplified macroevolutionary
models \cite{sole96,sole97}. Though these stylized models  do
reproduce the fractal-like structure of taxonomic trees, which
suggests that they successfully capture the essential ingredients
in the organization of biological taxa, they seldom give a
quantitatively satisfactory explanation of such regularities.
By now, however, there is little doubt that the power-law
distributions found in taxon abundance are a consequence of the
inherently complex mechanisms that drive biological
macroevolution, giving rise to speciation and, more generally,
originating new members at all the taxonomic levels.
Self-similarity and fractal features, in fact, have been
recognized as a clue to the underlying complexity in a large
class of dynamical systems \cite{mandelbrot97}. The fact that the
same kind of distributions is found in the taxonomy of languages
strongly suggests that language classification reflects, even at
its highest levels, the underlying evolutionary mechanisms.

In summary, we have shown that, at the highest levels of the
taxonomic classification of human languages, the frequency of
members containing a given number of languages decays as a
power-law over at least two decades. These systematic
regularities seem to discard the possibility that the
classification results from a baseless method. Moreover, they
imply that the frequency of taxon members containing a given
number of members from the successive taxon is also well
described by a power-law distribution. The same property is found
in the taxonomy of biological species, whose evolutionary origin
is firmly established. Along with the genetic evidence provided
by Cavalli-Sforza, this analogy supports an evolutionary basis
for Ruhlen's classification.

\section*{Acknowledgements}

Useful suggestions by G. Abramson, S. C. Manrubia, M. A.
Montemurro, and I. Samengo are gratefully acknowledged.

\end{document}